%% file: paper_arxiv.tex
\documentclass[conference]{IEEEtran}
\IEEEoverridecommandlockouts
\usepackage{cite}
\usepackage{amsmath,amssymb,amsthm}
\usepackage{bbm}
\usepackage{algpseudocode}
\usepackage{graphicx}
\usepackage{textcomp}
\usepackage{mathtools}
\usepackage{xcolor}
\usepackage{algorithm}
\usepackage{bm}

\def\BibTeX{{\rm B\kern-.05em{\sc i\kern-.025em b}\kern-.08em
    T\kern-.1667em\lower.7ex\hbox{E}\kern-.125emX}}
\usepackage{url}

\usepackage{tikz}
    \usetikzlibrary{shapes.arrows}
    
\usepackage{subfig}
\usepackage{pgfplots}
\pgfplotsset{
compat=1.3,
legend style={font=\footnotesize, fill opacity=0.7,  draw opacity=1, text opacity=1, draw=white!15!black, legend cell align=left, align=left}, 
width=6cm, 
height=6cm,
yminorticks=false,
xminorticks=false,
title style={font=\small},
tick style={color=black},
tick label style={font=\small},
grid style={line width=.1pt, draw=gray!20},
major grid style={line width=.1pt,draw=gray!20},
}
\usepgfplotslibrary{colorbrewer}
\usepgfplotslibrary{groupplots}
\usepgfplotslibrary{fillbetween}
\usetikzlibrary{decorations.pathreplacing}
\usetikzlibrary{plotmarks}
\usepackage[]{fancyhdr}
\usetikzlibrary{patterns}
\pgfplotsset{every tick label/.append style={font=\footnotesize}}

\usepackage{tikz}
\usetikzlibrary{shapes, arrows, positioning}

\usepackage{glossaries}
\input{acronyms.tex}

\DeclareMathOperator{\Mod}{mod}
\newcommand{\E}[1]{\mathbb{E}\left[ #1 \right]} 
\newcommand{\mc}[1]{\mathcal{#1}}   
\newcommand{\mb}[1]{\mathbf{#1}}    
\DeclareMathOperator*{\argmax}{arg\,max}    

\algtext*{EndWhile}
\algtext*{EndIf}
\algtext*{EndFor}

\def \fwidth{0.49\columnwidth}
\def \fheight {0.36\columnwidth}

\def \sfwidth{0.95\columnwidth}
\def \sfheight {0.5\columnwidth}
\def \boxside {0.3\columnwidth}
\def \tboxside {0.3\columnwidth}
\def \tboxheight {0.24\columnwidth}
\def\boxheight{0.24\columnwidth}

\definecolor{color0}{HTML}{00429D}
\definecolor{color1}{HTML}{844D99}
\definecolor{color2}{HTML}{C3608E}
\definecolor{color3}{HTML}{EF8078}
\definecolor{color4}{HTML}{FFB047}
\definecolor{darkslategray38}{RGB}{38,38,38}
\definecolor{darkblue}{HTML}{00429D}
\definecolor{darkgreen}{HTML}{005c00}
\definecolor{gold}{HTML}{D4AF37}
\definecolor{darkred}{HTML}{910000}

\newtheorem{theorem}{Theorem}

\makeatletter

\title{Eavesdropping on Goal-Oriented Communication: Timing Attacks and Countermeasures
}

\author{\IEEEauthorblockN{Federico Mason, Federico Chiariotti, Pietro Talli, and Andrea Zanella}
\IEEEauthorblockA{Department of Information Engineering, University of Padova, Via G. Gradenigo 6/B, 35131, Padua, Italy \\
Emails: federico.mason@unipd.it, federico.chiariotti@unipd.it, pietro.talli@phd.unipd.it, andrea.zanella@unipd.it}
\thanks{This project was funded under the National Recovery and Resilience Plan (NRRP), funded by the European Union NextGenerationEU Project as part of the ``RESTART'' partnership (PE0000001).}}

\begin{document}

\maketitle

\begin{abstract}
Goal-oriented communication is a new paradigm that considers the meaning of transmitted information to optimize communication.
One possible application is the remote monitoring of a process under communication costs: scheduling updates based on goal-oriented considerations can significantly reduce transmission frequency while maintaining high-quality tracking performance.
However, goal-oriented scheduling also opens a timing-based side-channel that an eavesdropper may exploit to obtain information about the state of the remote process, even if the content of updates is perfectly secure.
In this work, we study an eavesdropping attack against pull-based goal-oriented scheduling for the tracking of remote Markov processes.
We provide a theoretical framework for defining the effectiveness of the attack and of possible countermeasures, as well as a practical heuristic that can provide a balance between the performance gains offered by goal-oriented communication and the information leakage.
\end{abstract}


\glsresetall

\section{Introduction}
\label{sec:intro}

\begin{tikzpicture}[remember picture, overlay]
\node[draw, minimum width=4in, font=\small, text=black] at ([yshift=-1cm]current page.north)  {This paper has been accepted to the 8th Workshop on Age and Semantics of Information (ASoI 2025).};
\end{tikzpicture}

Over the past few years, the goal-oriented communication paradigm has attracted a significant amount of interest from the research community.
While more complex communication systems that go beyond the simple transmission of bits and consider the meaning and usefulness of the data were envisioned by Warren Weaver in his 1949 introduction to Shannon's theory of communication~\cite{shannon1949mathematical}, a practical implementation of these ideas requires powerful machine learning systems~\cite{gunduz2022beyond} and, therefore, has only recently become feasible.
The goal-oriented paradigm was initially applied to compression~\cite{deepjscoding}, but has successively been extended to packet scheduling that takes into account contextual and past information~\cite{fountoulakis2023goal}.

The initial research on goal-oriented communication has shown impressive performance in several use cases, which has led researchers to broaden their investigations to consider more practical aspects, such as the new paradigm's security against eavesdropping attacks~\cite{guo2024survey}.
In the current literature, the most common approach for such a goal is to directly adapt the learning architecture and training processes to include encryption~\cite{tung2023deep} and provide security properties as a secondary objective~\cite{liu2023semprotector}.
A complementary approach involves the exploitation of information theory~\cite{kung2018compressive} to provide more solid privacy guarantees~\cite{chen2024nearly}, under specific assumptions on the nature of the encoder and decoder.

However, there is a class of attacks against goal-oriented communication that has been mostly neglected so far: side-channel attacks that exploit the timing of messages instead of their content~\cite{van2015clock}.
This is particularly critical for \gls{iot} applications or other resource-constrained monitoring systems, where goal-oriented communication is used to adapt the frequency of updates as well as their content.
In these scenarios, timing attacks can leak information about the content of the updates even under perfect encryption.

The security of monitoring systems against side-channel attacks involves the concept of \emph{opacity}:
a system is opaque if an eavesdropper with limited observations is unable to estimate some restricted information~\cite{mazare2004decidability}, e.g., the identity of a client or whether the system enters a set of secret states.
The analysis of opacity has been extended to $K$-step observations~\cite{yin2017new} and even infinite sequences~\cite{saboori2011verification}, i.e., scenarios in which the eavesdropper has access to the whole observation history.
In information theoretic terms, opacity can be defined as the difference between the entropy of the belief distribution of the legitimate monitor and the eavesdropper~\cite{chen2017quantification}.

In this work, we analyze a goal-oriented communication system for monitoring a Markov source, modeling the information leakage associated with eavesdropping attacks.
We show that adapting the scheduling of state transmissions to balance cost with estimation accuracy reduces the opacity of the system by opening a side-channel that an eavesdropper can use to break system security.
Our objective is to maintain a certain level of privacy over the state for the past $D$ steps, even if the eavesdropper has full access to the complete history of transmission timings.
To the best of our knowledge, this work is the first to consider the security implications of timing attacks against goal-oriented communication.

Hence, our main contributions are the following:
\begin{itemize}
    \item we provide a rigorous model of timing attacks in goal-oriented communication, defining the information leakage as a function of the time for which privacy must be ensured; 
    \item we prove that finding a game theoretical equilibrium when both the legitimate agent and the eavesdropper are rational actors is a computationally hard problem;
    \item we propose a new algorithm, named \gls{ade}, that allows the legitimate agent to balance the trade-off between performance and estimation secrecy;
    \item we evaluate the effectiveness of the timing attack and of the defensive countermeasures by running multiple simulations in a simple estimation task.
\end{itemize}

The rest of the paper is organized as follows:
Sec.~\ref{sec:model} presents the goal-oriented communication model, drawing from results in our previous work~\cite{talli2024pragmatic}.
Sec.~\ref{sec:attack} presents the eavesdropping attack, game theoretical model, and heuristic countermeasure, while
Sec.~\ref{sec:results} discusses our simulation settings and results.
Finally, Sec.~\ref{sec:conc} concludes the paper and describes some possible avenues for future research.

\section{Goal-Oriented Communication Model}
\label{sec:model}

We consider a remote estimation system in which one node (Alice) can instantaneously observe the state $s_A(n)$ of a recurrent discrete-time Markov chain with a state space $\mc{S}$ and a transition matrix $\mb{P}$.
The initial distribution of the state is denoted by $\bm{\mu}_0$, and the steady-state distribution is denoted by $\bm{\mu}$.
The other node (Bob) must monitor the evolution of the process, but does not have direct access to the state, while both Alice and Bob have full knowledge of $\mb{P}$ and $\bm{\mu}_0$.

We consider a \emph{pull-based system}, in which at each time step, Bob must decide whether to ask Alice for information on the current state, incurring a communication cost $\beta$ but obtaining the real state, or estimating the current state only using the past information.
We denote Bob's binary communication decision for time step $n$ as $a(n)$ and his state estimate as $\hat{s}(n)$.
The objective function for Bob is given by the combination of the communication cost and his correctness in estimating the state:
\begin{equation}
    r(s,a,\hat{s})=\delta(s,\hat{s})\delta(a,0)+(1-\beta)\delta(a,1),
    \label{eq:effective_reward}
\end{equation}
where $\delta(m,n)$ is the Kronecker delta function, equal to $1$ if the two arguments are the same and $0$ otherwise.

We can then pose the problem as a \gls{pomdp}, in which Bob must use the available information to maximize the long-term reward with an exponential discount $\gamma$.
In this case, the scheduling policy ${\pi}(s,\Delta)$ depends only on the last received state $s$ and the time $\Delta$ since the last update.
We assume that the communication delay is lower than the Markov time step so that, whenever Alice transmits, Bob receives the state information instantaneously.
The \gls{mpi} scheme given in~\cite[Alg. 1]{talli2024pragmatic} can find the optimal scheduling policy in polynomial time over the state space size $|\mc{S}|$.

\begin{figure}[t!]
\input{tikz_fig/diagram}
\caption{The goal-oriented eavesdropping attack: Eve cannot decipher Alice's responses, but gets the timing signal $\tau$.}
\label{fig:diagram}
\end{figure}
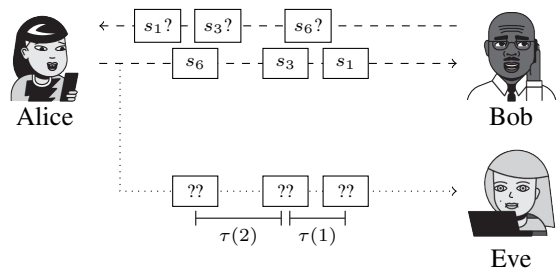

Since Bob uses goal-oriented scheduling, the timing of his requests depends on his estimate of the current state $s$.
Therefore, Bob’s strategy can be represented through the function 
\begin{equation}\label{eq:timing_policy}
    \sigma(s)=\inf\{\Delta \in \mathbb{N} : \pi(s,\Delta)=1\},
\end{equation}
returning the number of time steps that Bob waits after receiving state $s$ from Alice before asking for a new transmission.
If an eavesdropper (Eve) knows $\bm{\sigma}$, i.e., the mapping between the transmission intervals $\tau$ and the process state, as well as the source statistics $\mb{P}$ and $\bm{\mu}_0$, she can then use the timing of Bob's requests to gain information on the system itself.  
A diagram of the overall system is presented in Fig.~\ref{fig:diagram}.


From Eve's perspective, the system is a \gls{hmm}, where the timing signals $\tau$ are the observations that enable to compute of the \gls{map} estimate of the source.
Our formulation assumes that Eve only has access to the timing signals and does not consider the initial knowledge of Eve over the Markov source.
If the initial state distribution is low-entropy, the mixing time of the chain might be quite long, which leads to an edge case whose analysis is left to future work.

\section{Eavesdropping Attack and Countermeasure}
\label{sec:attack}

We consider a case where Alice and Bob want to prevent Eve from gaining information about the remote process within a maximum delay $D$.
Let $\bm{\phi}_E(n;d)$ denote Eve's belief over the state of the process at time $n-d$, after she has listened to the channel up to time $n$.
We can then define the system's information leakage at time step $n$ as
\begin{equation}
    L_E(n;D)=\argmax_{d\in\{0,\ldots,D\}} \left\{1-\frac{H\left(\bm{\phi}_E(n;d)\right)}{H(\bm{\mu})} \right\},
\label{eq:leakage}
\end{equation}
where $H(\cdot)$ is the Shannon information theoretic entropy~\cite{shannon1949mathematical}, defined as $H(\mb{p})=-\sum_{s\in\mc{S}}p(s)\log_2(p(s))$.
Notably, we have $L_E(n;D)=0$ if $\phi_E(n;D)=\bm{\mu}$, i.e., Eve does not have any additional information about the Markov source than the steady-state distribution.
Instead $L_E(n;D)=1$ if $\phi_E(n;D)=\delta(s,s_{n-d})$ for some $d$, i.e., Eve has perfect knowledge of the system state in at least one of the last $D$ steps.

\subsection{The Forward-Backward Algorithm}

Since Eve sees the system as an \gls{hmm}, the \gls{map} estimate of the system state can be computed through the forward-backward algorithm by combining forward probabilities, which only consider the past, with backward probabilities, which only consider the future. When estimating the state at time $m$ using information up to time $n>m$, forward probabilities are based on the observations from $0$ to $m$, while backward probabilities are based on those from $m$ to $n$.

Upon observing the $k$-th request from Bob, Eve can then compute the forward probability of any given state as
\begin{equation}
    f_k(s)=\sum_{s'\in\mc{S}}\left(\mb{P}^{\tau(k)}\right)_{s',s}\delta(\tau(k),\sigma(s'))f_{k-1}(s'),
\end{equation}
where $\mb{f}_0=\bm{\mu}_0$, $\tau(k)$ is the time observed by Eve between the $k-1$-th and the $k$-th transmission, and $\sigma(s)$ is the transmission policy as defined in~\eqref{eq:timing_policy}.
The backward probabilities are
\begin{equation}
    b_k(s;n)=\delta(\tau(k+1),\sigma(s))\sum_{s'\in\mc{S}}\left(\mb{P}^{\tau(k+1)}\right)_{s,s'}b_{k+1}(s';n),
\end{equation}
where $b_{K(n)}(s)=|\mc{S}|^{-1}\,\forall s\in\mc{S}$, as Eve has no information after this step, and $K(n)$ represents the index of the last transmission before time step $n$.
The \gls{map} estimate of the state when the $k$-th update is transmitted is then
\begin{equation}
    \phi_k(s;n)=\frac{f_k(s)b_k(s;n)}{\sum_{s'\in\mc{S}}f_k(s')b_k(s';n)}.
\end{equation}
Eve can also compute the \gls{map} estimate of the state of the Markov source $\ell$ steps after the $k$-th transmission step and $\tau(k+1)-\ell$ steps before the $k+1$-th transmission step as
\begin{equation}
\begin{aligned}
    \phi_{k,\ell}(s;n)=&\sum_{\mathclap{s',s''\in\mc{S}}}\phi_k(s';n)\phi_{k+1}(s'';n)\left(\mb{P}^{\ell}\right)_{s',s}\\
    &\times\left(\mb{P}^{\tau(k+1)-\ell}\right)_{s,s''}.
\end{aligned}
\end{equation}
Using the above formulas, Eve can compute $\bm{\phi}_E(n;d)$ for any value of $n$ and $d$.
We observe that the forward-backward algorithm's running time is $O(|\mc{S}|^2n)$, so it can be implemented with a relatively low energy cost: we can also limit the history duration to the mixing time of the Markov process to cap it.

\subsection{Game Theoretical Framework}

Since Eve is a purely adversarial attacker, whose goal is to obtain information on the Markov source or affect Bob's performance, we can model the system as a zero-sum one-sided partially observable stochastic game~\cite{horak2023solving}.
Bob aims to accurately estimate the Markov source without leaking information, while Eve's goal is the opposite, but her knowledge of the state is limited. 
The long-term reward for Bob is \begin{equation}
    R_B(D)=\E{\sum_{n=0}^\infty r(s(n),a(n),\hat{s}(n))-\varepsilon L_E(n;D)},
\end{equation}
where $\varepsilon>0$ is a parameter that can be used to adjust the relative importance of Bob's estimation accuracy with respect to information leakage.
Solutions based on the convexity property of the value function~\cite{horak2023solving} or on dividing the problem into sub-games with limited trajectories~\cite{delage2023hsvi} have recently been proposed, but their computational complexity increases exponentially with the state space size.

\begin{theorem}
    Finding the \gls{ne} to the zero-sum game between Bob and Eve has an exponentially growing computational time over the state space size $|\mc{S}|$.
\end{theorem}
\begin{IEEEproof}
    A classical result by Dantzig~\cite{dantzig1951proof} proves that a two-player zero-sum game with payoff matrix $\mb{M}$ is equivalent to the following linear programming problem:
    \begin{equation}
        \begin{aligned}
            \text{minimize }&\sum_i \mb{x}\quad          \text{such that }\mb{x}\geq0,\ \mb{M}\mb{x}=1.
        \end{aligned}
    \end{equation}
    Normalizing vector $\mb{x}$ returns the optimal mixed strategy for one of the players. In our case, the action space for Bob is equivalent to the possible policies he can adopt, which grows exponentially with the number of states $|\mc{S}|$. The length of $\mb{x}$ will also grow exponentially, making solving the game in polynomial time impossible.
\end{IEEEproof}

\subsection{Practical Countermeasure}

\begin{figure}[t]
\vspace{-0.3cm}
\begin{algorithm}[H]
\caption{\gls{ade}}
\label{alg:ppss}
\begin{algorithmic}[1]
\footnotesize

\Function{Schedule}{$s, \bm{\sigma},T,\mb{P}, \mb{f}, \mb{b}, \bm{\tau}, L_{\min}, L_{\max},\xi$}
\State $L_{\text{sem}}\gets L_E$ with $\tau(k)=\sigma(s)$
\State $L_{\text{per}}\gets L_E$ with $\tau(k)=T$
\If{$\xi=0$} \Comment{Goal-oriented scheduling active}
    \If {$L_{\text{sem}}\geq L_{\max}$} \Comment{Check privacy threshold}
    \State \Return{$T,1$}\Comment{Switch to PP}
    \Else
    \State \Return{$\sigma(s),0$}\Comment{Keep using MPI}
    \EndIf
\Else \Comment{Periodic scheduling active}
    \If {$L_{\text{per}}< L_{\min}$} \Comment{Check performance threshold}
    \State \Return{$\sigma(s),0$}\Comment{Switch to MPI}
    \Else
    \State \Return{$T,1$}\Comment{Keep using PP}
    \EndIf
\EndIf
\EndFunction
\end{algorithmic}
\end{algorithm}\vspace{-0.7cm}
\end{figure}

While finding an \gls{ne} is computationally intractable, we can design a simple heuristic policy that allows Bob to balance the performance advantages of goal-oriented communication and the system privacy.
We know that the optimal goal-oriented scheduling policy outperforms any periodic policy in terms of the expected reward, i.e., the trade-off between accuracy and transmission cost~\cite[Th. 2]{talli2024pragmatic}. However, it is also vulnerable to timing attacks, while a periodic policy does not leak any information, as we prove below.
\begin{theorem}
    If the proposed system is used to monitor a recurrent Markov chain, any periodic scheduling policy is perfectly private, i.e., information leakage tends to $0$ as $n$ increases for any finite value of $D$.
\end{theorem}
\begin{IEEEproof}
    Under a periodic policy with period $T$, we have $\sigma(s)=T\ \forall\, s \in \mc{S}$. Accordingly, the forward probabilities are $
        f_k(s)=\sum_{s'\in\mc{S}}\left(\mb{P}^T\right)_{s',s}f_{k-1}(s')$.
This is exactly equivalent to a blind update, and the same holds for the backward pass.
As timing provides no new information, Eve's belief tends to the steady-state distribution for any $n$ larger than the system mixing time, reducing leakage to $0$ as the window for the leakage calculation moves past the initial transient.
\end{IEEEproof}

We then exploit this in the proposed \acrfull{ade}: as Bob knows the timing signals, he can predict the information leakage for the next transmission interval. He can then switch from a goal-oriented to a periodic strategy when the leakage becomes higher than an upper threshold $L_{\max}$, then switch back to goal-oriented communication when the leakage goes below a lower threshold $L_{\min}$. This hysteresis pattern allows Bob to limit both the average and maximum leakage, while still exploiting goal-oriented communication at least in some time intervals.
The full \gls{ade} pseudocode is reported as Algorithm~\ref{alg:ppss}.

\section{Simulation Settings and Results}
\label{sec:results}

We consider a Markov chain with $|\mc{S}|=30$ states numbered from $1$ to $|\mc{S}|$, whose transition matrix $\mb{P}$ is defined as
\begin{equation}
P_{i,j} = 
    \begin{cases}
        \frac{1 + 2 g(i, \theta)}{3}, &   j = i \oplus 1 \wedge \Mod(i,4)\neq 2; \\
        \frac{2 - 2 g(i, \theta)}{3}, &   j = i \oplus 1 \wedge \Mod(i,4)= 2; \\
        \frac{1 - g(i, \theta)}{3}, & j\in\{i\oplus3,i\ominus2\}\wedge \Mod(i,4)\neq 2;\\
        \frac{2 + g(i, \theta)}{3}, & j\in\{i\oplus3,i\ominus2\}\wedge \Mod(i,4)= 2;\\
        0, & \text{otherwise,}
    \end{cases}
    \label{eq:transition_prob}
\end{equation}
where $\oplus$ and $\ominus$ represent modulo $|\mc{S}|$ addition and subtraction,  $\Mod(m,n)$ is the integer modulo function, $\theta \in \mathbb{R}^+$ is a parameter named \emph{density decay}, and $g(i, \theta)$ is defined as
\begin{equation}
    g(i, \theta) =\left(| 2i - |\mathcal{S}| |\right)^{\theta}|\mathcal{S}|^{-\theta}.
    \label{eq:transition_std}
\end{equation}
There is a high probability of going from state $i$ to state $i\oplus1$, and a lower probability of going to either state $i\ominus2$ or $i\oplus3$. In one state every $4$, this distribution is reversed, with a high probability of avoiding the next state. The structure of the matrix is designed to provide a single tuning parameter to control the predictability of the source evolution, while the reversed states are included to avoid trivial edge cases.

We note that $g(i, \theta)$ is one at the extremes of the state space (i.e., for $i=0$ and $i=|S|$) and progressively decreases when moving towards middle states. This implies that states farther from the middle tend to have more deterministic transitions to the next state, while the allowed transitions have similar probabilities for states closer to the middle. Moreover, the randomness of the state transitions can be tuned through $\theta \in \mathbb{R}^+$. 
As $\theta \rightarrow \infty$, $g(i, \theta)$ tends to zero and most states will have uniform (i.e., unpredictable) transition probabilities to neighboring states; conversely, as $\theta \rightarrow 0$, $g(i, \theta)$ tends to $1$ and most states will have deterministic (and, hence, fully predictable) transitions.

We generate multiple configurations for the communication system by varying both the density decay $\theta \in [1, 2^7]$ and the transmission cost $\beta \in [0.2, 2]$. 
We then compute the optimal goal-oriented scheduling given by the \gls{mpi} algorithm~\cite{talli2024pragmatic} for each configuration.
In doing so, we set $T_{\text{max}} = 10$ as the maximum interval between two consecutive transmissions, i.e., the maximum value that $\tau$ can take.

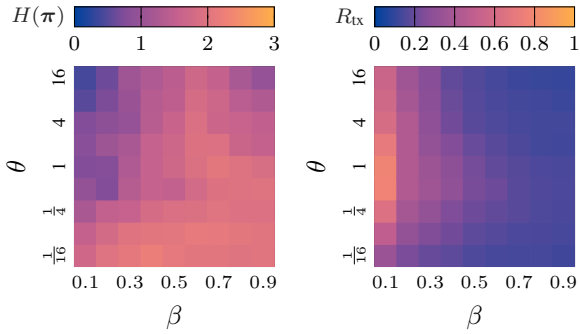
\begin{figure}[t!]
    \centering
    \subfloat[Entropy of the timing signal.\label{fig:entropy}]
    {\input{./tikz_fig/entropy_effective.tex}}
    \subfloat[Transmission probability.\label{fig:transmission}]
    {\input{./tikz_fig/transmission_effective.tex}}
    \caption{Characterization of the optimal scheduling policy as a function of the density decay $\theta$ and the transmission cost $\beta$.}
    \label{fig:markov_heatmaps}
\end{figure}

Fig.~\ref{fig:entropy} represents the entropy $H(\bm{\tau})$ of the distribution of the timing signals associated with each state, which is an indicator of the information that the scheduling policy $\bm{\sigma}$ provides to Eveon the state $s$. 
A periodic policy would have zero entropy, as the inter-transmission time is fixed, while a policy that selects a different value of $\sigma(s)$ for each state would have an entropy equal to $\log_2(|\mc{S}|)$. In general, $H(\bm{\tau})$ decreases as $\beta \rightarrow 0$: if the transmission cost is lower, transmissions are more common, and more states are associated to the same inter-transmission time.
We observe a similar result as $\theta$ increases: in this case, the entropy $H(\bm{\tau})$ is reduced because state transitions are less predictable, and the overall performance depends much more on the transmission probability.
As Fig.~\ref{fig:transmission} shows, the transmission probability decreases as $\beta$ increases.
On the other hand, when $\theta$ grows, the evolution of the process is characterized by a strong randomness, which makes it inconvenient to trigger new transmissions, except for the states with a larger $g(i, \theta)$.

\begin{figure}[t!]
    \centering
    \input{./tikz_fig/leakage_time.tex}
    \caption{Information leakage during a single episode, with $\beta=1$, $\theta=32$, and $D=5$. The ADE thresholds $L_{\min}$ and $L_{\max}$ are marked as dashed lines.}
    \label{fig:leak_vs_time}
\end{figure}
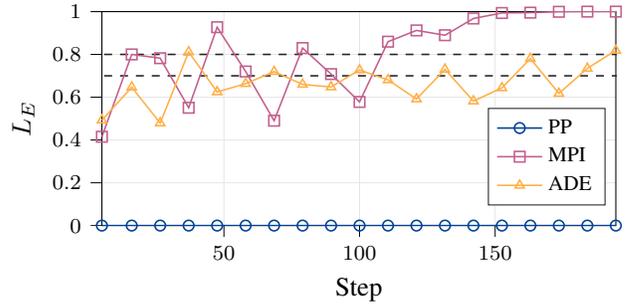

In the following, we evaluate the performance of the \gls{ade} heuristic against two possible benchmarks: the optimal scheduling policy obtained via the purely goal-oriented \gls{mpi} algorithm and a \gls{pp}, for which the scheduling decisions are agnostic to the state observations. 
In particular, the inter-transmission period of \gls{pp} was tuned to maximize the long-term reward, while \gls{ade} was configured to maintain the leakage value between $L_{\min}=0.4$ and $L_{\max}=0.6$. 
To evaluate the different strategies, we run a total of $N_{\text{ep}}=10$ episodes for each configuration, considering $N_{\text{step}}=200$ steps per episode, and analyze the results in terms of the expected reward $r$ and leakage $L_{E}$. 
We expect the latter to be correlated to $H(\bm{\tau})$: if timing signals can take more values, Eve's observation space will be larger and more informative. 

We can have an overview of how the different strategies behave in Fig.~\ref{fig:leak_vs_time}, which reports the information leakage during an episode in a system with $\beta=1$, $\theta=32$, and $D=5$.
The leakage of the purely goal-oriented \gls{mpi} algorithm oscillates, as the process moves through more and less predictable regions of the state space. However, $L_E$ is often close to $0.9$, allowing Eve to correctly guess the state about two thirds of the time. Conversely, \gls{pp} does not provide any information to Eve, who can only guess the state about $5\%$ of the time, as her knowledge is based only on the steady-state probability of the Markov process. However, the overall reward when using \gls{mpi} increases by about $40\%$. 
Finally, the \gls{ade} algorithm ensures that the leakage oscillates between $L_{\min}$ and $L_{\max}$, resulting in a compromise between the two approaches: Eve can correctly guess the state about one third of the time, and the overall reward is about $20\%$ higher than when using \gls{pp}.

\begin{figure}[t!]
    \centering
    \subfloat[Leakage.\label{fig:leak_per}]
    {\input{./tikz_fig/leakage_periodic.tex}}
    \subfloat[Reward.\label{fig:rew_per}]
    {\input{./tikz_fig/reward_periodic.tex}}\vspace{-0.25cm}\\
    \subfloat[Bob's accuracy.\label{fig:bob_per}]
    {\input{./tikz_fig/acc_bob_periodic.tex}}
    \subfloat[Eve's accuracy.\label{fig:eve_per}]
    {\input{./tikz_fig/acc_eve_periodic.tex}}    
    \caption{\gls{pp} performance as a function of the density decay $\theta$ and the communication cost $\beta$, with $D=5$.}
    \label{fig:pp_heatmaps}
\end{figure}

\begin{figure}[t!]
    \subfloat[Leakage.\label{fig:leak_eff}]
    {\input{./tikz_fig/leakage_effective.tex}}
    \subfloat[Reward.\label{fig:rew_eff}]
    {\input{./tikz_fig/reward_effective.tex}}\vspace{-0.25cm}\\
    \subfloat[Bob's accuracy.\label{fig:bob_eff}]
    {\input{./tikz_fig/acc_bob_effective.tex}}
    \subfloat[Eve's accuracy.\label{fig:eve_eff}]
    {\input{./tikz_fig/acc_eve_effective.tex}}
    \caption{\gls{mpi} performance as a function of the density decay $\theta$ and the communication cost $\beta$, with $D=5$.}
    \label{fig:mpi_heatmaps}
\end{figure}

We now consider the performance of the policies with different values of the communication cost $\beta$ and density decay $\theta$.
We analyze Bob's accuracy in the estimation task, i.e., the expected frequency $\eta_B=\E{\delta(s,\hat{s})}$ of Bob correctly estimating the state, as well as Eve's, denoted by $\eta_E=\E{\delta\left(s_n,\argmax_{s'\in\mc{S}}\left[\phi_E(n+D,D)\right](s')\right)}$.
This setup gives Eve an advantage, as she can wait up to $D$ steps before estimating the state, while Bob must do so without the benefit of hindsight.
Fig.~\ref{fig:leak_per} clearly shows that $L_{E} \approx 0$ for all system configurations when Bob uses \gls{pp}: as predicted, periodic communication is fully opaque to timing attacks.
However, the expected reward, shown in Fig.~\ref{fig:rew_per}, degrades in the case of high communication cost ($\beta \rightarrow 2$) and stochastic transitions ($\theta \gg 1$). Bob's ability to estimate the state of the Markov source degrades as transmissions become less frequent due to the higher cost, as we can observe in Fig.~\ref{fig:bob_per}.

The purely goal-oriented \gls{mpi} algorithm can improve the system reward in these conditions by approximately $25\%$, as shown in Fig.~\ref{fig:rew_eff}-\subref*{fig:bob_eff}. Although \gls{mpi} tends to transmit more than \gls{pp}, Bob's accuracy does not significantly decline as $\beta$ increases. The gain over \gls{pp} in terms of Bob's accuracy reaches $50\%$ when $\beta\rightarrow2$.
On the other hand, \gls{mpi} significantly reduces the system secrecy, as shown in Fig.~\ref{fig:leak_eff}: the information leakage is close to $0.8$ for all configurations except for those with very high transmission cost and density decay values. 
Critically, Eve is able to correctly decode the status of the monitored process almost as often as Bob, as Fig.~\ref{fig:eve_eff} shows, highlighting the vulnerability of \gls{mpi} to timing attacks. 
The most vulnerable configurations are those characterized by more predictable transitions and low transmission costs. 
In these scenarios, the \gls{mpi} algorithm leads to scheduling decisions with a high entropy and a high transmission rate, as shown in Fig.~\ref{fig:markov_heatmaps}.

Finally, the proposed \gls{ade} heuristic is able to improve secrecy in all configurations, as shown in Fig.~\ref{fig:leak_heu}, by limiting the leakage value to $L_{\max}$.
Fig.~\ref{fig:rew_heu} shows that this also causes a degradation of the expected reward, especially in the case of Markov sources with a low $\theta$ and a high transmission cost. The reward obtained by \gls{ade} is in between the performance of \gls{mpi} and \gls{pp}, with a performance gain of approximately $10\%$ over \gls{pp}.
These results are confirmed by Fig.~\ref{fig:bob_heu}-\subref*{fig:eve_heu}: Eve's accuracy decreases much more than Bob's, with a mean leakage of $0.45$, confirming that \gls{ade} can effectively control the trade-off between performance and opacity.

\begin{figure}[t!]
    \subfloat[Leakage.\label{fig:leak_heu}]
    {\input{./tikz_fig/leakage_heuristic.tex}}
    \subfloat[Reward.\label{fig:rew_heu}]
    {\input{./tikz_fig/reward_heuristic.tex}} \vspace{-0.25cm}\\ 
    \subfloat[Bob's accuracy.\label{fig:bob_heu}]
    {\input{./tikz_fig/acc_bob_heuristic.tex}}
    \subfloat[Eve's accuracy.\label{fig:eve_heu}]
    {\input{./tikz_fig/acc_eve_heuristic.tex}}
    \caption{\gls{ade} performance as a function of the density decay $\theta$ and the communication cost $\beta$, with $D=5$.}
    \label{fig:ade_heatmaps}
\end{figure}

\begin{figure}[t!]
    \centering
\subfloat{\input{tikz_fig/delay_legend}}\\         \setcounter{subfigure}{0}
\subfloat[Information leakage.\label{fig:delay_leak}]
{\input{./tikz_fig/delay_leakage.tex}}
\subfloat[Reward.\label{fig:delay_rew}]
{\input{./tikz_fig/delay_reward.tex}}
    \caption{Expected leakage and reward as a function of the maximum delay $D$, with $\beta=1$ and $\theta=32$.}
    \label{fig:delay_impact}
\end{figure}
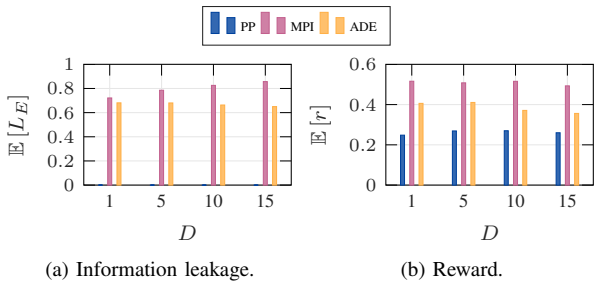

We finally analyze the impact of the maximum delay on performance in Fig.~\ref{fig:delay_impact}, setting $D\in\{1, 5, 10, 15\}$. Interestingly, while the leakage of \gls{mpi} grows as $D$ increases, \gls{ade} tends to make more conservative choices in this case, and its leakage actually decreases. However, this choice results in a slightly lower reward, as \gls{ade} must switch to \gls{pp} more often and for longer periods, while the reward is unchanged for \gls{pp} and \gls{mpi}.

\section{Conclusion and Future Work}
\label{sec:conc}

This work presents an important issue of goal-oriented communication scheduling strategies in remote monitoring systems: while this approach has significant performance benefits in terms of the trade-off between the estimation accuracy and transmission cost, it is also vulnerable to eavesdropping.
Timing attacks are viable even under information-theoretic secrecy, as they only rely on the presence of a message instead of its content.
Our results show that, while heuristic mitigation strategies are possible, finding an optimal policy under game theoretic rationality is a computationally hard problem.

As our study is the first to analyze timing attacks against goal-oriented communication, there are many possible avenues of future work. Firstly, the expansion of the game theoretic model may lead to more efficient heuristics.
Hence, it will be interesting to consider reinforcement learning solutions, which have similar properties to the proposed algorithms and can be deployed in more complex real-world scenarios.
Finally, extending the model to push-based scenarios, in which Alice independently decides when to send an update, is an another appealing research possibility.

\bibliographystyle{IEEEtran}
\bibliography{biblio.bib}

\end{document}

%% file: acronyms.tex
\newacronym{iot}{IoT}{Internet of Things}
\newacronym{ml}{ML}{machine learning}
\newacronym{dl}{DL}{Deep Learning}
\newacronym{marl}{MARL}{multi-agent reinforcement learning}
\newacronym{rl}{RL}{reinforcement learning}
\newacronym{decpomdp}{Dec-POMDP}{Decentralized Partially Observable Markov Decision Process }
\newacronym{acnomdp}{ACNO-MDP}{Action-Contingent Noiselessly Observable MDP}
\newacronym{pomdp}{POMDP}{Partially Observable Markov Decision Process}
\newacronym{lidar}{LIDAR}{Laser Imaging Detection and Ranging}
\newacronym{uav}{UAV}{Unmanned Aerial Vehicle}
\newacronym{dqn}{DQN}{Deep Q-Network}
\newacronym{dnn}{DNN}{deep neural network}
\newacronym{dial}{DIAL}{Differentiable Inter-Agent Learning}
\newacronym{api}{API}{Alternate Policy Iteration}
\newacronym{mdp}{MDP}{Markov Decision Process}
\newacronym{fov}{FoV}{Field of View}
\newacronym{cnn}{CNN}{Convolutional Neural Network}
\newacronym{nn}{NN}{neural network}
\newacronym{ddql}{DDQL}{Distributed Deep Q-Learning}
\newacronym{pdf}{PDF}{Probability Density Function}
\newacronym{ndpomdp}{ND-POMDP}{Networked Distributed Partially Observable Markov Decision Process}
\newacronym{ncs}{NCS}{Networked Control System}
\newacronym{cps}{CPS}{Cyber-Physical System}
\newacronym{radam}{RAdam}{Rectified Adam}
\newacronym{cdf}{CDF}{cumulative distribution function}
\newacronym{mpc}{MPC}{Model Predictive Control}
\newacronym{rv}{rv}{Random Variable}
\newacronym{qoe}{QoE}{Quality of Experience}
\newacronym{tlc}{TLC}{Telecommunications}
\newacronym{cml}{CML}{communications for machine learning}
\newacronym{mlc}{MLC}{machine learning for communications}
\newacronym{drl}{DRL}{Deep Reinforcement Learning}
\newacronym{rf}{RF}{Radio Frequency}
\newacronym{urllc}{URLLC}{Ultra-Reliable and Low-Latency Communications}
\newacronym{fl}{FL}{federated learning}
\newacronym{kpi}{KPI}{Key Performance Indicators}
\newacronym{mec}{MEC}{Mobile Edge Computing}
\newacronym{ei}{EI}{Edge Intelligence}
\newacronym{bs}{BS}{base station}
\newacronym{sdn}{SDN}{Software Defined Networking}
\newacronym{mimo}{MIMO}{Multiple-Input Multiple-Output}
\newacronym{gp}{GP}{Gaussian Process}
\newacronym{iiot}{IIoT}{Industrial Internet of Things}
\newacronym{csi}{CSI}{Channel State Information}
\newacronym{sgd}{SGD}{Stochastic Gradient Descent}
\newacronym{iid}{i.i.d.}{independent and identically distributed}
\newacronym{ofdm}{OFDM}{Orthogonal Frequency Division Multiplexing}
\newacronym{los}{LOS}{Line-of-Sight}
\newacronym{nlos}{NLOS}{Non-Line-of-Sight}
\newacronym{snr}{SNR}{Signal to Noise Ratio}
\newacronym{rb}{RB}{Resource Block}
\newacronym{6g}{6G}{sixth generation}
\newacronym{ai}{AI}{artificial intelligence}
\newacronym{sfl}{SFL}{Synchronous Federated Learning}
\newacronym{frfl}{FRFL}{Fixed Rate Federated Learning}
\newacronym{pgm}{PGM}{Probabilistic Graphical Model}
\newacronym{hmm}{HMM}{Hidden Markov Model}
\newacronym{elbo}{ELBO}{Evidence Lower Bound}
\newacronym{pmf}{PMF}{Probability Mass Function}
\newacronym{smab}{SMAB}{Stochastic Multi-Armed Bandit}
\newacronym{mab}{MAB}{multi-armed bandit}
\newacronym{mc}{MC}{Monte Carlo}
\newacronym{is}{IS}{Importance Sampling}
\newacronym{dms}{DMS}{discrete memoryless source}
\newacronym{ucb}{UCB}{upper confidence bound}
\newacronym{ser}{SER}{Symbol Error Rate}
\newacronym{sc}{SC}{Semantic Communications}
\newacronym{voi}{VoI}{Value of Information}
\newacronym{nlp}{NLP}{natural language processing}
\newacronym{ts}{TS}{Thompson Sampling}
\newacronym{cmab}{CMAB}{contextual multi-armed bandit}
\newacronym{rccmab}{RC-CMAB}{rate-constrained \gls{cmab}}
\newacronym{rcmab}{R-CMAB}{remote \gls{cmab}}
\newacronym{ib}{IB}{information bottleneck}
\newacronym{merl}{MERL}{maximum entropy reinforcement learning}
\newacronym{pac}{PAC}{Probably Approximately Correct}
\newacronym{aoi}{AoI}{Age of Information}
\newacronym{aoii}{AoII}{Age of Incorrect Information}
\newacronym{uoi}{UoI}{Urgency of Information}
\newacronym{vqvae}{VQ-VAE}{Vector Quantized Variational Autoencoder}
\newacronym{vae}{VAE}{Variational Autoencoder}
\newacronym{psnr}{PSNR}{Peak Signal to Noise Ratio}
\newacronym{mse}{MSE}{Mean Square Error}
\newacronym{lstm}{LSTM}{Long Short-Term Memory}
\newacronym{iomdp}{IOMDP}{Intermittently Observable Markov Decision Process}
\newacronym{pi}{PI}{Policy Iteration}
\newacronym{mpi}{MPI}{Modified Policy Iteration}
\newacronym{iql}{IQL}{Intermittent Q-Learning}
\newacronym{nc}{NC}{Network Core}
\newacronym{pql}{PQL}{Peekaboo Q-Learning}
\newacronym{eff_com}{EC}{Effective Communication}
\newacronym{ibr}{IBR}{Iterated Best Response}
\newacronym{ne}{NE}{Nash Equilibrium}
\newacronym{map}{MAP}{maximum a posteriori}
\newacronym{ade}{ADE}{Alternating Defense from Eavesdropping}
\newacronym{pp}{PP}{Periodic Policy}

%% file: tikz_fig/diagram.tex
\begin{tikzpicture}[auto]
\node (alice) at (0,0) {\includegraphics[width=1.2cm]{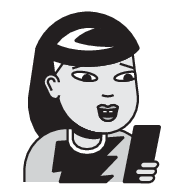}};
\node[name=atxt] at (0,-0.8) {{Alice}};

\node (bob) at (6.2,0) {\includegraphics[width=1.2cm]{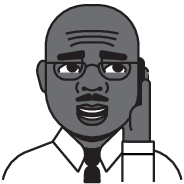}};
\node[name=btxt] at (6.2,-0.8) {Bob};

\node (eve) at (6.2,-1.8) {\includegraphics[width=1.2cm]{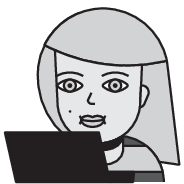}};
\node[name=rtxt] at (6.2,-2.7) {Eve};

\draw[<-,dashed]([yshift=0.4cm]alice.east) to ([yshift=0.4cm]bob.west);
\draw[->,dashed]([yshift=-0.1cm]alice.east) to ([yshift=-0.1cm]bob.west);

\draw[->,dotted](1,-0.1) |- (eve.west);

\node[rectangle,draw,fill=white,minimum height=0.4cm,minimum width=0.6cm] at (4,-0.1) {\scriptsize $s_1$};

\node[rectangle,draw,fill=white,minimum height=0.4cm,minimum width=0.6cm] at (3.2,-0.1) {\scriptsize $s_3$};

\node[rectangle,draw,fill=white,minimum height=0.4cm,minimum width=0.6cm] at (2,-0.1) {\scriptsize $s_6$};

\node[rectangle,draw,fill=white,minimum height=0.4cm,minimum width=0.6cm] at (3.5,0.4) {\scriptsize $s_6$?};

\node[rectangle,draw,fill=white,minimum height=0.4cm,minimum width=0.6cm] at (2.3,0.4) {\scriptsize $s_3$?};

\node[rectangle,draw,fill=white,minimum height=0.4cm,minimum width=0.6cm] at (1.5,0.4) {\scriptsize $s_1$?};

\node (p1)[rectangle,draw,fill=white,minimum height=0.4cm,minimum width=0.6cm] at (2,-1.8) {\scriptsize ??};

\node (p2) [rectangle,draw,fill=white,minimum height=0.4cm,minimum width=0.6cm] at (3.2,-1.8) {\scriptsize ??};

\node (p3)[rectangle,draw,fill=white,minimum height=0.4cm,minimum width=0.6cm] at (4,-1.8) {\scriptsize ??};

\draw[|-|] ([yshift=-0.15cm]p3.south) -- node[midway,below]{\scriptsize $\tau(1)$}([xshift=0.05cm,yshift=-0.15cm]p2.south);

\draw[|-|] ([yshift=-0.15cm]p1.south) -- node[midway,below]{\scriptsize $\tau(2)$}([xshift=-0.05cm,yshift=-0.15cm]p2.south);



\end{tikzpicture}

%% file: tikz_fig/entropy_effective.tex
\begin{tikzpicture}
    \begin{axis}[
    width=\tboxside,
    height=\tboxheight,
    tick align=outside,
    ytick pos = left,
    xtick pos = bottom,
    scale only axis,
    name=lin,
    xlabel=$\beta$,
    ylabel=$\log_2(\theta)$,
    mesh/cols=8,
    mesh/rows=10,
    yticklabel style={rotate=90,font={\scriptsize}},
    xticklabel style={font={\scriptsize}},
    xmin=0.1,
    xmax=2.1,
    ymin=0.5,
    ymax=8.5,
    ytick={1,2,3,4,5,6,7,8},
    yticklabels={0,1,2,3,4,5,6,7},
    xtick={0.2,0.6,1,1.4,1.8},
    point meta min=0,
    point meta max=3,
colormap={mymap}{[1pt]
rgb(0pt)=(0, 0.258823529411765, 0.615686274509804);
rgb(1pt)=(0.517647058823530, 0.301960784313725, 0.600000000000000);
rgb(2pt)=(0.764705882352941, 0.376470588235294, 0.556862745098039);
rgb(3pt)=(0.937254901960784, 0.501960784313726, 0.470588235294118);
rgb(4pt)=(1, 0.690196078431373, 0.278431372549020);
 },
    colorbar horizontal,
    colorbar style={
    at={(0,1.3)},
    height=0.1*\pgfkeysvalueof{/pgfplots/parent axis width},
ylabel style={rotate=-90,font={\footnotesize\color{white!15!black}}},ylabel={$H(\bm{\tau})$},
                    yticklabel style={
                        /pgf/number format/fixed,
                        /pgf/number format/precision=2
                }}
]
    \addplot[matrix plot*, point meta=explicit] file {./tikz_fig/data/entropy.dat};
\end{axis}

\end{tikzpicture}

%% file: tikz_fig/transmission_effective.tex
\begin{tikzpicture}
    \begin{axis}[
    width=\tboxside,
    height=\tboxheight,
    tick align=outside,
    ytick pos = left,
    xtick pos = bottom,
    scale only axis,
    name=lin,
    xlabel=$\beta$,
    ylabel=$\log_2(\theta)$,
    mesh/cols=8,
    mesh/rows=10,
    yticklabel style={rotate=90,font={\scriptsize}},
    xticklabel style={font={\scriptsize}},
    xmin=0.1,
    xmax=2.1,
    ymin=0.5,
    ymax=8.5,
    ytick={1,2,3,4,5,6,7,8},
    yticklabels={0,1,2,3,4,5,6,7},
    xtick={0.2,0.6,1,1.4,1.8},
    point meta min=0,
    point meta max=1,
colormap={mymap}{[1pt]
rgb(0pt)=(0, 0.258823529411765, 0.615686274509804);
rgb(1pt)=(0.517647058823530, 0.301960784313725, 0.600000000000000);
rgb(2pt)=(0.764705882352941, 0.376470588235294, 0.556862745098039);
rgb(3pt)=(0.937254901960784, 0.501960784313726, 0.470588235294118);
rgb(4pt)=(1, 0.690196078431373, 0.278431372549020);
 },
    colorbar horizontal,
    colorbar style={
    at={(0,1.3)},
    height=0.1*\pgfkeysvalueof{/pgfplots/parent axis width},
ylabel style={rotate=-90,font={\footnotesize\color{white!15!black}}},ylabel={$R_{\text{tx}}$},
                    yticklabel style={
                        /pgf/number format/fixed,
                        /pgf/number format/precision=2
                }}
]
    \addplot[matrix plot*,point meta=explicit] file {./tikz_fig/data/transmission.dat};
\end{axis}

\end{tikzpicture}

%% file: tikz_fig/leakage_time.tex
\begin{tikzpicture}

\begin{axis}[%
width=\sfwidth,
height=\sfheight,
tick align=outside,
xlabel={Step},
xmajorgrids,
xmin=5, 
xmax=194.5,
tick pos=left,
legend style={legend cell align=left, fill opacity=1, draw opacity=1, text opacity=1, legend columns=3, align=left, draw=white!15!black, font=\footnotesize, at={(0.965, 0.1)}, anchor=south east},
ylabel={$L_{E}$},
ymajorgrids,
ymin=0, ymax=1,
]

\addplot [semithick, darkslategray38, dashed, forget plot]
table {
0   0.4
200 0.4
};

\addplot [semithick, darkslategray38, dashed, forget plot]
table {
0   0.6
200 0.6
};

\addplot [semithick, color0, mark=o]
table {%
5   0.0001
16  0.0001
26.5   0.0001
37  0.0001
47.5   0.000
58  0.0001
68.5  0.0001
79 0.0001
89.5   0.0001
100 0.0001
110.5   0.0001
121 0.0001
131.5   0.0001
142  0.0001
152.5  0.0001
163  0.0001
173.5   0.0001
184  0.0001
194.5  0.0001
};
\addlegendentry{PP};

, , , , , , , , , 0.999999771339994, 0.9999999980474958, 0.9999999999512785, 0.9999999999991968, 0.9999999999999791, 0.9999999999999997, 1.0, 1.0, 1.0, 1.0

\addplot [semithick, color2, mark=square]
table {%
5   0.3599942507983917
16  0.46192201961279494
26.5    0.565887866403214
37  0.6966921165223464
47.5    0.8656900183717703
58  0.9326288026723326
68.5    0.9053334707281481
79  0.9083708485227479
89.5    0.8266665194428521
100 0.7654384364237533
110.5   0.8843509106272931
121 0.9149762257574668
131.5   0.8775902764047763
142 0.8607325818049034
152.5   0.9153130395891174
163 0.9253413156931792
173.5   0.6482759949690943
184  0.7012422688550638
194.5   0.7757889801954767
};
\addlegendentry{MPI};

\addplot [semithick, color4, mark=triangle]
table {%
5   0.2528888180755532
16  0.46229928843585105
26.5    0.5778379343506177
37  0.41106993108905737
47.5    0.41974234586940806
58  0.539436513135603
68.5    0.41529830483872293
79  0.5006879978671713
89.5    0.5138498963486806
100 0.44083839270728614
110.5   0.34145773828240794
121 0.4254462908626847
131.5   0.5458878129577474
142 0.44548055226838185
152.5   0.5258410069293195
163 0.4019707358195884
173.5   0.5493718301961605
184 0.5487979456076348
194.5   0.3745820591476491
};
\addlegendentry{ADE};

\end{axis}

\end{tikzpicture}

%% file: tikz_fig/leakage_periodic.tex
\begin{tikzpicture}
    \begin{axis}[
    width=\boxside,
    height=\boxheight,
    tick align=outside,
    ytick pos = left,
    xtick pos = bottom,
    scale only axis,
    name=lin,
    xlabel=$\beta$,
    ylabel=$\log_2(\theta)$,
    mesh/cols=8,
    mesh/rows=10,
    yticklabel style={rotate=90,font={\scriptsize}},
    xticklabel style={font={\scriptsize}},
    xmin=0.1,
    xmax=2.1,
    ymin=0.5,
    ymax=8.5,
    ytick={1,2,3,4,5,6,7,8},
    yticklabels={0,1,2,3,4,5,6,7},
    xtick={0.2,0.6,1,1.4,1.8},
    point meta min=0,
    point meta max=1,
colormap={mymap}{[1pt]
rgb(0pt)=(0, 0.258823529411765, 0.615686274509804);
rgb(1pt)=(0.517647058823530, 0.301960784313725, 0.600000000000000);
rgb(2pt)=(0.764705882352941, 0.376470588235294, 0.556862745098039);
rgb(3pt)=(0.937254901960784, 0.501960784313726, 0.470588235294118);
rgb(4pt)=(1, 0.690196078431373, 0.278431372549020);
 },
    colorbar horizontal,
    colorbar style={
    at={(0,1.35)},
    height=0.1*\pgfkeysvalueof{/pgfplots/parent axis width},
ylabel style={rotate=-90,font={\footnotesize\color{white!15!black}}},ylabel={$\E{L_E}$},
                    yticklabel style={
                        /pgf/number format/fixed,
                        /pgf/number format/precision=2
                }}
]
    \addplot[matrix plot*, point meta=explicit] file {./tikz_fig/periodic_data/leak_vs_decay_vs_cost.dat};
\end{axis}

\end{tikzpicture}

%% file: tikz_fig/reward_periodic.tex
\begin{tikzpicture}
    \begin{axis}[
    width=\boxside,
    height=\boxheight,
    scale only axis,
    tick align=outside,
    ytick pos = left,
    xtick pos = bottom,
    scale only axis,
    name=lin,
    xlabel=$\beta$,
    ylabel=$\log_2(\theta)$,
    mesh/cols=8,
    mesh/rows=10,
    yticklabel style={rotate=90,font={\scriptsize}},
    xticklabel style={font={\scriptsize}},
    xmin=0.1,
    xmax=2.1,
    ymin=0.5,
    ymax=8.5,
    ytick={1,2,3,4,5,6,7,8},
    yticklabels={0,1,2,3,4,5,6,7},
    xtick={0.2,0.6,1,1.4,1.8},
    point meta min=0,
    point meta max=1,
colormap={mymap}{[1pt]
rgb(0pt)=(0, 0.258823529411765, 0.615686274509804);
rgb(1pt)=(0.517647058823530, 0.301960784313725, 0.600000000000000);
rgb(2pt)=(0.764705882352941, 0.376470588235294, 0.556862745098039);
rgb(3pt)=(0.937254901960784, 0.501960784313726, 0.470588235294118);
rgb(4pt)=(1, 0.690196078431373, 0.278431372549020);
 },
    colorbar horizontal,
    colorbar style={
    at={(0,1.35)},
    height=0.1*\pgfkeysvalueof{/pgfplots/parent axis width},
ylabel style={rotate=-90,font={\footnotesize\color{white!15!black}}},ylabel={$\E{r}$},
                    yticklabel style={
                        /pgf/number format/fixed,
                        /pgf/number format/precision=2
                }}
]
    \addplot[matrix plot*, point meta=explicit] file {./tikz_fig/periodic_data/total_reward_vs_decay_vs_cost.dat};
\end{axis}

\end{tikzpicture}

%% file: tikz_fig/acc_bob_periodic.tex
\begin{tikzpicture}
    \begin{axis}[
    width=\boxside,
    height=\boxheight,
    tick align=outside,
    ytick pos = left,
    xtick pos = bottom,
    scale only axis,
    name=lin,
    xlabel=$\beta$,
    ylabel=$\log_2(\theta)$,
    mesh/cols=8,
    mesh/rows=10,
    yticklabel style={rotate=90,font={\scriptsize}},
    xticklabel style={font={\scriptsize}},
    xmin=0.1,
    xmax=2.1,
    ymin=0.5,
    ymax=8.5,
    ytick={1,2,3,4,5,6,7,8},
    yticklabels={0,1,2,3,4,5,6,7},
    xtick={0.2,0.6,1,1.4,1.8},
    point meta min=0,
    point meta max=1,
colormap={mymap}{[1pt]
rgb(0pt)=(0, 0.258823529411765, 0.615686274509804);
rgb(1pt)=(0.517647058823530, 0.301960784313725, 0.600000000000000);
rgb(2pt)=(0.764705882352941, 0.376470588235294, 0.556862745098039);
rgb(3pt)=(0.937254901960784, 0.501960784313726, 0.470588235294118);
rgb(4pt)=(1, 0.690196078431373, 0.278431372549020);
 },
    colorbar horizontal,
    colorbar style={
    at={(0,1.35)},
    height=0.1*\pgfkeysvalueof{/pgfplots/parent axis width},
ylabel style={rotate=-90,font={\footnotesize\color{white!15!black}}},ylabel={$\eta_B$},
                    yticklabel style={
                        /pgf/number format/fixed,
                        /pgf/number format/precision=2
                }}
]
    \addplot[matrix plot*, point meta=explicit] file {./tikz_fig/periodic_data/bob_accuracy_vs_decay_vs_cost.dat};
\end{axis}

\end{tikzpicture}

%% file: tikz_fig/acc_eve_periodic.tex
\begin{tikzpicture}
    \begin{axis}[
    width=\boxside,
    height=\boxheight,
    tick align=outside,
    ytick pos = left,
    xtick pos = bottom,
    scale only axis,
    name=lin,
    xlabel=$\beta$,
    ylabel=$\log_2(\theta)$,
    mesh/cols=8,
    mesh/rows=10,
    yticklabel style={rotate=90,font={\scriptsize}},
    xticklabel style={font={\scriptsize}},
    xmin=0.1,
    xmax=2.1,
    ymin=0.5,
    ymax=8.5,
    ytick={1,2,3,4,5,6,7,8},
    yticklabels={0,1,2,3,4,5,6,7},
    xtick={0.2,0.6,1,1.4,1.8},
    point meta min=0,
    point meta max=1,
colormap={mymap}{[1pt]
rgb(0pt)=(0, 0.258823529411765, 0.615686274509804);
rgb(1pt)=(0.517647058823530, 0.301960784313725, 0.600000000000000);
rgb(2pt)=(0.764705882352941, 0.376470588235294, 0.556862745098039);
rgb(3pt)=(0.937254901960784, 0.501960784313726, 0.470588235294118);
rgb(4pt)=(1, 0.690196078431373, 0.278431372549020);
 },
    colorbar horizontal,
    colorbar style={
    at={(0,1.35)},
    height=0.1*\pgfkeysvalueof{/pgfplots/parent axis width},
ylabel style={rotate=-90,font={\footnotesize\color{white!15!black}}},ylabel={$\eta_E$},
                    yticklabel style={
                        /pgf/number format/fixed,
                        /pgf/number format/precision=2
                }}
]
    \addplot[matrix plot*, point meta=explicit] file {./tikz_fig/periodic_data/eve_accuracy_vs_decay_vs_cost.dat};
\end{axis}

\end{tikzpicture}

%% file: tikz_fig/leakage_effective.tex
\begin{tikzpicture}
    \begin{axis}[
    width=\boxside,
    height=\boxheight,
    tick align=outside,
    ytick pos = left,
    xtick pos = bottom,
    scale only axis,
    name=lin,
    xlabel=$\beta$,
    ylabel=$\log_2(\theta)$,
    mesh/cols=8,
    mesh/rows=10,
    yticklabel style={rotate=90,font={\scriptsize}},
    xticklabel style={font={\scriptsize}},
    xmin=0.1,
    xmax=2.1,
    ymin=0.5,
    ymax=8.5,
    ytick={1,2,3,4,5,6,7,8},
    yticklabels={0,1,2,3,4,5,6,7},
    xtick={0.2,0.6,1,1.4,1.8},
    point meta min=0,
    point meta max=1,
colormap={mymap}{[1pt]
rgb(0pt)=(0, 0.258823529411765, 0.615686274509804);
rgb(1pt)=(0.517647058823530, 0.301960784313725, 0.600000000000000);
rgb(2pt)=(0.764705882352941, 0.376470588235294, 0.556862745098039);
rgb(3pt)=(0.937254901960784, 0.501960784313726, 0.470588235294118);
rgb(4pt)=(1, 0.690196078431373, 0.278431372549020);
 },
    colorbar horizontal,
    colorbar style={
    at={(0,1.35)},
    height=0.1*\pgfkeysvalueof{/pgfplots/parent axis width},
ylabel style={rotate=-90,font={\footnotesize\color{white!15!black}}},ylabel={$\E{L_E}$},
                    yticklabel style={
                        /pgf/number format/fixed,
                        /pgf/number format/precision=2
                }}
]
    \addplot[matrix plot*, point meta=explicit] file {./tikz_fig/effective_data/leak_vs_decay_vs_cost.dat};
\end{axis}

\end{tikzpicture}

%% file: tikz_fig/reward_effective.tex
\begin{tikzpicture}
    \begin{axis}[
    width=\boxside,
    height=\boxheight,
    tick align=outside,
    ytick pos = left,
    xtick pos = bottom,
    scale only axis,
    name=lin,
    xlabel=$\beta$,
    ylabel=$\log_2(\theta)$,
    mesh/cols=8,
    mesh/rows=10,
    yticklabel style={rotate=90,font={\scriptsize}},
    xticklabel style={font={\scriptsize}},
    xmin=0.1,
    xmax=2.1,
    ymin=0.5,
    ymax=8.5,
    ytick={1,2,3,4,5,6,7,8},
    yticklabels={0,1,2,3,4,5,6,7},
    xtick={0.2,0.6,1.0,1.4,1.8},
    point meta min=0,
    point meta max=1,
colormap={mymap}{[1pt]
rgb(0pt)=(0, 0.258823529411765, 0.615686274509804);
rgb(1pt)=(0.517647058823530, 0.301960784313725, 0.600000000000000);
rgb(2pt)=(0.764705882352941, 0.376470588235294, 0.556862745098039);
rgb(3pt)=(0.937254901960784, 0.501960784313726, 0.470588235294118);
rgb(4pt)=(1, 0.690196078431373, 0.278431372549020);
 },
    colorbar horizontal,
    colorbar style={
    at={(0,1.35)},
    height=0.1*\pgfkeysvalueof{/pgfplots/parent axis width},
ylabel style={rotate=-90,font={\footnotesize\color{white!15!black}}},ylabel={$\E{r}$},
                    yticklabel style={
                        /pgf/number format/fixed,
                        /pgf/number format/precision=2
                }}
]
    \addplot[matrix plot*, point meta=explicit] file {./tikz_fig/effective_data/total_reward_vs_decay_vs_cost.dat};
\end{axis}

\end{tikzpicture}

%% file: tikz_fig/acc_bob_effective.tex
\begin{tikzpicture}
    \begin{axis}[
    width=\boxside,
    height=\boxheight,
    tick align=outside,
    ytick pos = left,
    xtick pos = bottom,
    scale only axis,
    name=lin,
    xlabel=$\beta$,
    ylabel=$\log_2(\theta)$,
    mesh/cols=8,
    mesh/rows=10,
    yticklabel style={rotate=90,font={\scriptsize}},
    xticklabel style={font={\scriptsize}},
    xmin=0.1,
    xmax=2.1,
    ymin=0.5,
    ymax=8.5,
    ytick={1,2,3,4,5,6,7,8},
    yticklabels={0,1,2,3,4,5,6,7},
    xtick={0.2,0.6,1,1.4,1.8},
    point meta min=0,
    point meta max=1,
colormap={mymap}{[1pt]
rgb(0pt)=(0, 0.258823529411765, 0.615686274509804);
rgb(1pt)=(0.517647058823530, 0.301960784313725, 0.600000000000000);
rgb(2pt)=(0.764705882352941, 0.376470588235294, 0.556862745098039);
rgb(3pt)=(0.937254901960784, 0.501960784313726, 0.470588235294118);
rgb(4pt)=(1, 0.690196078431373, 0.278431372549020);
 },
    colorbar horizontal,
    colorbar style={
    at={(0,1.35)},
    height=0.1*\pgfkeysvalueof{/pgfplots/parent axis width},
ylabel style={rotate=-90,font={\footnotesize\color{white!15!black}}},ylabel={$\eta_B$},
                    yticklabel style={
                        /pgf/number format/fixed,
                        /pgf/number format/precision=2
                }}
]
    \addplot[matrix plot*, point meta=explicit] file {./tikz_fig/effective_data/bob_accuracy_vs_decay_vs_cost.dat};
\end{axis}

\end{tikzpicture}

%% file: tikz_fig/acc_eve_effective.tex
\begin{tikzpicture}
    \begin{axis}[
    width=\boxside,
    height=\boxheight,
    tick align=outside,
    ytick pos = left,
    xtick pos = bottom,
    scale only axis,
    name=lin,
    xlabel=$\beta$,
    ylabel=$\log_2(\theta)$,
    mesh/cols=8,
    mesh/rows=10,
    yticklabel style={rotate=90,font={\scriptsize}},
    xticklabel style={font={\scriptsize}},
    xmin=0.1,
    xmax=2.1,
    ymin=0.5,
    ymax=8.5,
    ytick={1,2,3,4,5,6,7,8},
    yticklabels={0,1,2,3,4,5,6,7},
    xtick={0.2,0.6,1,1.4,1.8},
    point meta min=0,
    point meta max=1,
colormap={mymap}{[1pt]
rgb(0pt)=(0, 0.258823529411765, 0.615686274509804);
rgb(1pt)=(0.517647058823530, 0.301960784313725, 0.600000000000000);
rgb(2pt)=(0.764705882352941, 0.376470588235294, 0.556862745098039);
rgb(3pt)=(0.937254901960784, 0.501960784313726, 0.470588235294118);
rgb(4pt)=(1, 0.690196078431373, 0.278431372549020);
 },
    colorbar horizontal,
    colorbar style={
    at={(0,1.35)},
    height=0.1*\pgfkeysvalueof{/pgfplots/parent axis width},
ylabel style={rotate=-90,font={\footnotesize\color{white!15!black}}},ylabel={$\eta_E$},
                    yticklabel style={
                        /pgf/number format/fixed,
                        /pgf/number format/precision=2
                }}
]
    \addplot[matrix plot*, point meta=explicit] file {./tikz_fig/effective_data/eve_accuracy_vs_decay_vs_cost.dat};
\end{axis}

\end{tikzpicture}

%% file: tikz_fig/leakage_heuristic.tex
\begin{tikzpicture}
    \begin{axis}[
    width=\boxside,
    height=\boxheight,
    tick align=outside,
    ytick pos = left,
    xtick pos = bottom,
    scale only axis,
    name=lin,
    xlabel=$\beta$,
    ylabel=$\log_2(\theta)$,
    mesh/cols=8,
    mesh/rows=10,
    yticklabel style={rotate=90,font={\scriptsize}},
    xticklabel style={font={\scriptsize}},
    xmin=0.1,
    xmax=2.1,
    ymin=0.5,
    ymax=8.5,
    ytick={1,2,3,4,5,6,7,8},
    yticklabels={0,1,2,3,4,5,6,7},
    xtick={0.2,0.6,1,1.4,1.8},
    point meta min=0,
    point meta max=1,
colormap={mymap}{[1pt]
rgb(0pt)=(0, 0.258823529411765, 0.615686274509804);
rgb(1pt)=(0.517647058823530, 0.301960784313725, 0.600000000000000);
rgb(2pt)=(0.764705882352941, 0.376470588235294, 0.556862745098039);
rgb(3pt)=(0.937254901960784, 0.501960784313726, 0.470588235294118);
rgb(4pt)=(1, 0.690196078431373, 0.278431372549020);
 },
    colorbar horizontal,
    colorbar style={
    at={(0,1.35)},
    height=0.1*\pgfkeysvalueof{/pgfplots/parent axis width},
ylabel style={rotate=-90,font={\footnotesize\color{white!15!black}}},ylabel={$\E{L_E}$},
                    yticklabel style={
                        /pgf/number format/fixed,
                        /pgf/number format/precision=2
                }}
]
    \addplot[matrix plot*, point meta=explicit] file {./tikz_fig/heuristic_data/leak_vs_decay_vs_cost.dat};
\end{axis}

\end{tikzpicture}

%% file: tikz_fig/reward_heuristic.tex
\begin{tikzpicture}
    \begin{axis}[
    width=\boxside,
    height=\boxheight,
    tick align=outside,
    ytick pos = left,
    xtick pos = bottom,
    scale only axis,
    name=lin,
    xlabel=$\beta$,
    ylabel=$\log_2(\theta)$,
    mesh/cols=8,
    mesh/rows=10,
    yticklabel style={rotate=90,font={\scriptsize}},
    xticklabel style={font={\scriptsize}},
    xmin=0.1,
    xmax=2.1,
    ymin=0.5,
    ymax=8.5,
    ytick={1,2,3,4,5,6,7,8},
    yticklabels={0,1,2,3,4,5,6,7},
    xtick={0.2,0.6,1,1.4,1.8},
    point meta min=0,
    point meta max=1,
colormap={mymap}{[1pt]
rgb(0pt)=(0, 0.258823529411765, 0.615686274509804);
rgb(1pt)=(0.517647058823530, 0.301960784313725, 0.600000000000000);
rgb(2pt)=(0.764705882352941, 0.376470588235294, 0.556862745098039);
rgb(3pt)=(0.937254901960784, 0.501960784313726, 0.470588235294118);
rgb(4pt)=(1, 0.690196078431373, 0.278431372549020);
 },
    colorbar horizontal,
    colorbar style={
    at={(0,1.35)},
    height=0.1*\pgfkeysvalueof{/pgfplots/parent axis width},
ylabel style={rotate=-90,font={\footnotesize\color{white!15!black}}},ylabel={$\E{r}$},
                    yticklabel style={
                        /pgf/number format/fixed,
                        /pgf/number format/precision=2
                }}
]
    \addplot[matrix plot*, point meta=explicit] file {./tikz_fig/heuristic_data/total_reward_vs_decay_vs_cost.dat};
\end{axis}

\end{tikzpicture}

%% file: tikz_fig/acc_bob_heuristic.tex
\begin{tikzpicture}
    \begin{axis}[
    width=\boxside,
    height=\boxheight,
    tick align=outside,
    ytick pos = left,
    xtick pos = bottom,
    scale only axis,
    name=lin,
    xlabel=$\beta$,
    ylabel=$\log_2(\theta)$,
    mesh/cols=8,
    mesh/rows=10,
    yticklabel style={rotate=90,font={\scriptsize}},
    xticklabel style={font={\scriptsize}},
    xmin=0.1,
    xmax=2.1,
    ymin=0.5,
    ymax=8.5,
    ytick={1,2,3,4,5,6,7,8},
    yticklabels={0,1,2,3,4,5,6,7},
    xtick={0.2,0.6,1,1.4,1.8},
    point meta min=0,
    point meta max=1,
colormap={mymap}{[1pt]
rgb(0pt)=(0, 0.258823529411765, 0.615686274509804);
rgb(1pt)=(0.517647058823530, 0.301960784313725, 0.600000000000000);
rgb(2pt)=(0.764705882352941, 0.376470588235294, 0.556862745098039);
rgb(3pt)=(0.937254901960784, 0.501960784313726, 0.470588235294118);
rgb(4pt)=(1, 0.690196078431373, 0.278431372549020);
 },
    colorbar horizontal,
    colorbar style={
    at={(0,1.35)},
    height=0.1*\pgfkeysvalueof{/pgfplots/parent axis width},
ylabel style={rotate=-90,font={\footnotesize\color{white!15!black}}},ylabel={$\eta_B$},
                    yticklabel style={
                        /pgf/number format/fixed,
                        /pgf/number format/precision=2
                }}
]
    \addplot[matrix plot*, point meta=explicit] file {./tikz_fig/heuristic_data/bob_accuracy_vs_decay_vs_cost.dat};
\end{axis}

\end{tikzpicture}

%% file: tikz_fig/acc_eve_heuristic.tex
\begin{tikzpicture}
    \begin{axis}[
    width=\boxside,
    height=\boxheight,
    tick align=outside,
    ytick pos = left,
    xtick pos = bottom,
    scale only axis,
    name=lin,
    xlabel=$\beta$,
    ylabel=$\log_2(\theta)$,
    mesh/cols=8,
    mesh/rows=10,
    yticklabel style={rotate=90,font={\scriptsize}},
    xticklabel style={font={\scriptsize}},
    xmin=0.1,
    xmax=2.1,
    ymin=0.5,
    ymax=8.5,
    ytick={1,2,3,4,5,6,7,8},
    yticklabels={0,1,2,3,4,5,6,7},
    xtick={0.2,0.6,1,1.4,1.8},
    point meta min=0,
    point meta max=1,
colormap={mymap}{[1pt]
rgb(0pt)=(0, 0.258823529411765, 0.615686274509804);
rgb(1pt)=(0.517647058823530, 0.301960784313725, 0.600000000000000);
rgb(2pt)=(0.764705882352941, 0.376470588235294, 0.556862745098039);
rgb(3pt)=(0.937254901960784, 0.501960784313726, 0.470588235294118);
rgb(4pt)=(1, 0.690196078431373, 0.278431372549020);
 },
    colorbar horizontal,
    colorbar style={
    at={(0,1.35)},
    height=0.1*\pgfkeysvalueof{/pgfplots/parent axis width},
ylabel style={rotate=-90,font={\footnotesize\color{white!15!black}}},ylabel={$\eta_E$},
                    yticklabel style={
                        /pgf/number format/fixed,
                        /pgf/number format/precision=2
                }}
]
    \addplot[matrix plot*, point meta=explicit] file {./tikz_fig/heuristic_data/eve_accuracy_vs_decay_vs_cost.dat};
\end{axis}

\end{tikzpicture}

%% file: tikz_fig/delay_legend.tex
\begin{tikzpicture}

\pgfplotstableread{
D   pp  mpi shade
0   0   0   0
}\loadedtable;

\begin{axis}[%
width=0cm,
height=0cm,
ybar,
scale only axis,
tick align=inside,
bar width=1.5pt,
legend style={legend cell align=left, fill opacity=1, draw opacity=1, text opacity=1, legend columns=5, align=left, draw=white!15!black, font=\tiny, at={(0.5, 0.02)}, anchor=south},
xmin=0,
xmax=0,
xlabel={$D$},
ymin=0,
ymax=0,
ylabel={$\E{L_E}$},
axis background/.style={fill=white}
]

    \addplot[style={color0,fill={white!20!color0}}] table[x=D, y=pp] {\loadedtable};
    \addlegendentry{PP};
    \addplot[style={color2,fill={white!20!color2}}] table[x=D, y=mpi] {\loadedtable}; 
    \addlegendentry{MPI};
    \addplot[style={color4,fill={white!20!color4}}] table[x=D, y=shade] {\loadedtable}; 
    \addlegendentry{ADE};

    \end{axis}
\end{tikzpicture}

%% file: tikz_fig/delay_leakage.tex
\begin{tikzpicture}

\pgfplotstableread{
D   pp  mpi ade   stdpp   stdmpi  stdade
1   0   0.7774422940117902	0.49683256462131564  0 0.103   0.02
2   0   0.8090845580872892	0.46004920081808826  0 0.121   0.019
3   0   0.8600719626985953 0.4137862425442684   0    0.089   0.034
4   0   0.878541704340438	0.35580485369147175   0   0.064   0.033
}\loadedtable;

\begin{axis}[%
width=\fwidth,
height=\fheight,
ybar,
tick pos=left,
tick align=outside,
bar width=3pt,
xlabel style={font=\footnotesize\color{white!15!black}},
ylabel style={font=\footnotesize\color{white!15!black}},
tick label style={font=\scriptsize\color{white!15!black}},
xmajorgrids,
ymajorgrids,
xtick={1,2,3,4},
xticklabels={1,5,10,15},
xmin=0.5,
xmax=4.5,
xlabel={$D$},
ymin=0,
ymax=1,
ylabel={$\E{L_E}$},
axis background/.style={fill=white}
]

    \addplot[style={color0,fill={white!20!color0}}, error bars/.cd, error bar style={color=black}, error mark options={}, y dir=both, y explicit] table[x=D, y=pp, y error=stdpp] {\loadedtable};
    \addplot[style={color2,fill={white!20!color2}}, error bars/.cd, error bar style={color=black}, error mark options={}, y dir=both, y explicit] table[x=D, y=mpi, y error=stdmpi] {\loadedtable}; 
    \addplot[style={color4,fill={white!20!color4}}, error bars/.cd, error bar style={color=black}, error mark options={}, y dir=both, y explicit] table[x=D, y=ade, y error=stdade] {\loadedtable}; 

    \end{axis}
\end{tikzpicture}

%% file: tikz_fig/delay_reward.tex
\begin{tikzpicture}

\pgfplotstableread{
D   pp  mpi ade   stdpp   stdmpi   stdade
1   0.441  0.6155  0.553    0.105   0.037   0.072
2   0.4325  0.626  0.532    0.069   0.023   0.036
3   0.4285  0.621  0.5345   0.077   0.039   0.055
4   0.4065  0.6195  0.5085  0.073   0.054   0.09
}\loadedtable;

\begin{axis}[%
width=\fwidth,
height=\fheight,
ybar,
tick pos=left,
tick align=outside,
bar width=3pt,
xlabel style={font=\footnotesize\color{white!15!black}},
ylabel style={font=\footnotesize\color{white!15!black}},
tick label style={font=\scriptsize\color{white!15!black}},
xmajorgrids,
ymajorgrids,
xtick={1,2,3,4},
xticklabels={1,5,10,15},
xmin=0.5,
xmax=4.5,
xlabel={$D$},
ymin=0.3,
ymax=0.7,
ylabel={$\E{r}$},
axis background/.style={fill=white}
]

    \addplot[style={color0,fill={white!20!color0}}, error bars/.cd, error bar style={color=black}, error mark options={}, y dir=both, y explicit] table[x=D, y=pp, y error=stdpp] {\loadedtable};
    \addplot[style={color2,fill={white!20!color2}}, error bars/.cd, error bar style={color=black}, error mark options={}, y dir=both, y explicit] table[x=D, y=mpi, y error=stdmpi] {\loadedtable}; 
    \addplot[style={color4,fill={white!20!color4}}, error bars/.cd, error bar style={color=black}, error mark options={}, y dir=both, y explicit] table[x=D, y=ade, y error=stdade] {\loadedtable}; 

    \end{axis}
\end{tikzpicture}